\pdfoutput=1

\documentclass[11pt]{article}

\usepackage[preprint]{acl}

\usepackage{times}
\usepackage{latexsym}

\usepackage[T1]{fontenc}

\usepackage[utf8]{inputenc}

\usepackage{microtype}

\usepackage{inconsolata}

\usepackage{algorithm}
\usepackage{algorithmic}
\usepackage{xcolor}
\usepackage{pifont}
\usepackage{natbib}
\usepackage{caption}
\usepackage{graphicx}
\usepackage{booktabs}
\usepackage{multirow}
\usepackage{amsmath} 
\usepackage{amsfonts}
\usepackage{amsthm}
\usepackage{bm}
\usepackage{tcolorbox}
\usepackage{enumitem}
\usepackage{subcaption}
\usepackage{makecell}
\usepackage{subcaption}
\newcommand{\nc}[1]{{\textbf{\color{black} [\textit{\color{green}\underline{\color{purple} NEED CITE!}}]}}}
\newcommand{\nr}[1]{{\textbf{\color{black} [\textit{\color{green}\underline{\color{purple} NEED REF!}}]}}}

\usepackage[normalem]{ulem}
\useunder{\uline}{\ul}{}

\title{Similarity = Value? Consultation Value Assessment and Alignment for Personalized Search}

\author{
 \textbf{Weicong Qin\textsuperscript{1}\thanks{This work was done during an internship at Lenovo
 Inc.}},
 \textbf{Yi Xu\textsuperscript{1}\footnotemark[1]},
 \textbf{Weijie Yu\textsuperscript{2}\thanks{Corresponding author.}},
 \textbf{Teng Shi\textsuperscript{1}},
 \textbf{Chenglei Shen\textsuperscript{1}\footnotemark[1]},
\\
 \textbf{Ming He\textsuperscript{3}},
 \textbf{Jianping Fan\textsuperscript{3}},
 \textbf{Xiao Zhang\textsuperscript{1}},
 \textbf{Jun Xu\textsuperscript{1}},
\\
 \textsuperscript{1}Gaoling School of Artificial Intelligence, Renmin University of China, China\\
 \textsuperscript{2}University of International Business and Economics, China\\
 \textsuperscript{3}AI Lab at Lenovo Research, Lenovo Group Limited, China\\
\texttt{{yu@uibe.edu.cn}}
\\
}
\begin{document}
\maketitle
\begin{abstract}

Personalized search systems in e-commerce platforms increasingly involve user interactions with AI assistants, where users consult about products, usage scenarios, and more. Leveraging consultation to personalize search services is trending. Existing methods typically rely on semantic similarity to align historical consultations with current queries due to the absence of `value' labels, but we observe that semantic similarity alone often fails to capture the true value of consultation for personalization.
To address this, we propose a consultation value assessment framework that evaluates historical consultations from three novel perspectives: (1) Scenario Scope Value, (2) Posterior Action Value, and (3) Time Decay Value. Based on this, we introduce VAPS, a value-aware personalized search model that selectively incorporates high-value consultations through a consultation–user action interaction module and an explicit objective that aligns consultations with user actions.
Experiments on both public and commercial datasets show that VAPS consistently outperforms baselines in both retrieval and ranking tasks. Codes are available at \url{https://anonymous.4open.science/r/VAPS-to-go}.
\end{abstract}

\section{Introduction}
\label{sec:intro}
\begin{figure}[!t]
        \centering
\includegraphics[width=0.47\textwidth]{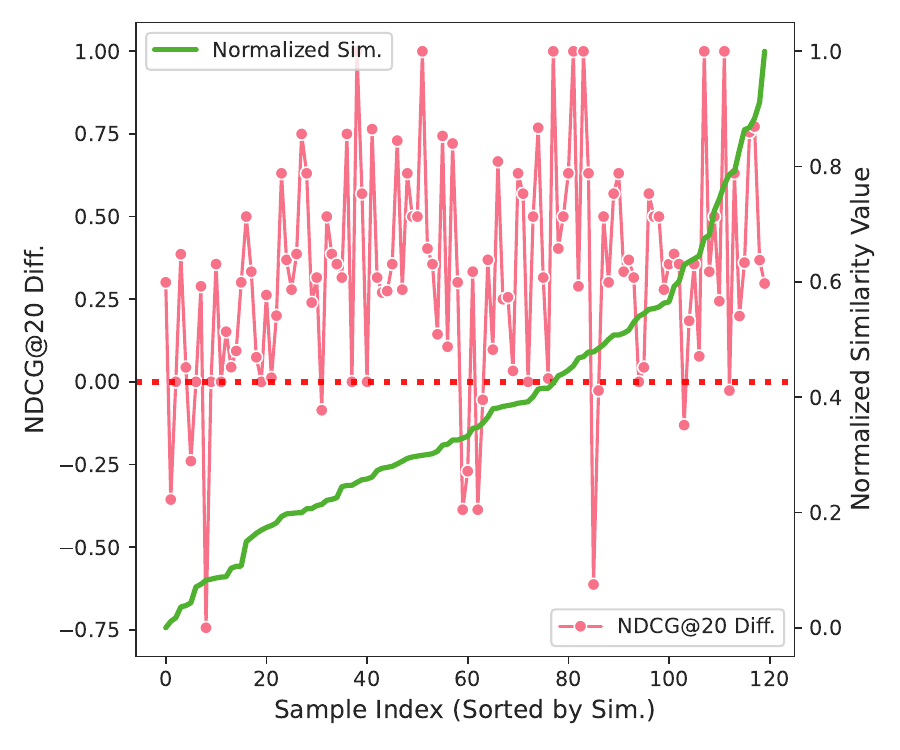}
    \caption{
    NDCG@20 difference with/without consultations vs. normalized semantic similarity, using MAPS~\cite{maps} on Amazon dataset. Semantic similarity does not consistently reflect the value of consultations for improving search.
    }
  \label{fig:intro-diff}
\end{figure}


Personalized product search~\citep{ai2019zero, shi2024unisar,shi2025unified} is key for e-commerce and search engines, delivering tailored results by leveraging user data. As AI assistants become more common in e-commerce, users increasingly rely on bots for product queries, comparisons, and availability checks. These consultation logs provide valuable insights into user intent beyond explicit searches, helping improve personalization and search accuracy~\cite{maps}.



Recent studies can be broadly divided into two directions. The first mainly extracts user-specific preferences from user interactions (across multiple scenarios) to personalize search results~\cite{bi2020transformer}, but lacks exploration of needs in user consultations. The second direction, represented by Motivation-Aware Personalized Search model (MAPS)~\cite{maps}, while utilizing user interaction data, primarily attempts to align current search queries with consultation histories via semantic similarity to tap into search needs within them. Due to the absence of `value' labels, these methods assume that semantically related past consultations inherently benefit current searches. However, as shown in Fig.~\ref{fig:intro-diff}, our analysis of MAPS' search performance differences and consultation-query semantic similarity reveals: semantic similarity cannot fully reliably reflect search utility. Consultations with high similarity scores may harm ranking performance, while some with lower similarity can improve it, even though an overall positive correlation exists. This suggests that semantic alignment is an inadequate proxy for identifying valuable consultations.

\begin{figure*}[!t]
\centering
\includegraphics[width=1\textwidth]{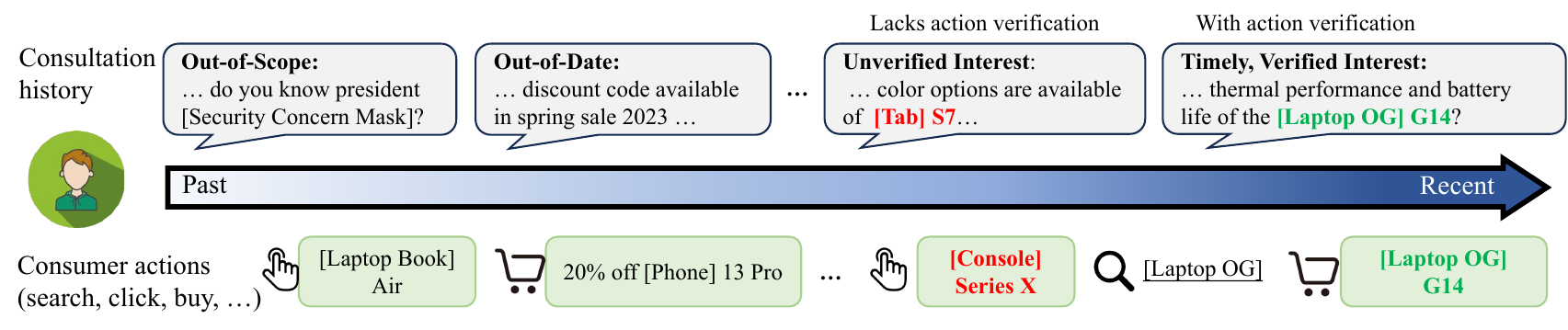}
  \caption{Illustration of consumer consultation patterns and action verification in E-commerce platform. Yellow background indicates outdated information; red text means no action verified for the interest; green text shows verified needs with actions.}
  \label{fig:intro-motivation}
\end{figure*}

Inspired by the consumer journey~\cite{hamilton2019consumer} in e-commerce, we argue that the value of a historical consultation should be assessed in the context of the user's evolving goals and actions. 
There are distinct consumer consultation-action patterns for users as illustrated in Fig.~\ref{fig:intro-motivation}. (1) Out-of-scope consultations, such as those related to politics, are regarded as noise. (2) Out-of-date offers such as the 2023 spring sale discount code are no longer useful. (3) Interest in products like the color options of the ``[Pad Tab] S7'' lacks action verification as no related actions follow. In contrast, inquiries about the  ``[Laptop OG] G14'' are accompanied by related actions, showing verified interest.

We identify three key factors affecting a consultation's value for personalized search: (1) Scenario Scope Value: whether the consultation pertains to the platform's product or service domain; (2) Time Decay Value: whether the consultation has expired, and (3)  Posterior Action Value: whether the consultation is followed by user relevant actions (e.g., click, search, purchase) indicating its influence.
For example, a timely inquiry that leads to a purchase conveys stronger search-relevant intent than an outdated or out-of-scope inquiry. However, it still faces the challenge of missing explicit labels.

To operationalize these insights and address the challenge, we propose a Consultation Value Assessment Framework that quantifies the value of historical consultations with empirical function-based assessment. Based on this, we introduce VAPS (Value-Aware  Consultation-Enhanced Personalized Search) model that selectively incorporates high-value consultations with a consultation–action interaction module and explicitly aligns consultations with consumer action signals. We further design tailored loss functions to ensure our methods capturing the value-rich consultation-action signals. Extensive
experiments on real and synthetic data show VAPS significantly outperforms baselines in both retrieval and ranking tasks.

To summarize, our contributions are as follows:
\begin{itemize}[leftmargin=*]
\item We pioneer the value assessment of user-AI consultations in personalized search, exposing flaws in current semantic-similarity-only method.
\item We propose a consultation value assessment framework with three novel dimensions (Scenario Scope, Time Decay, and Posterior Action) tailored for personalized search, addressing gaps in prior work.
\item We introduce VAPS, a value-aware consultation-enhanced personalized search model designed to align with value assessment signals
\item  Extensive experiments across retrieval and ranking stages, conducted on both real-world commercial and synthetic datasets, demonstrate that VAPS outperforms existing personalized search methods, traditional/conversational retrieval approaches, and multi-scenario models.
\end{itemize}

\section{Related Works}

Personalized search provides relevant items based on user queries~\citep{shi2024unisar}. Traditional methods like BM25~\cite{bm25} focus on word frequency, while dense retrieval (e.g., BGE-M3~\cite{chen2024bge}) uses embeddings. Conversational methods like CHIQ~\cite{mo2024chiq} improve accuracy using search history, but lack personalization.

Recent work includes QEM~\cite{ai2019zero} and DREM~\cite{ai2019explainable} for query-item similarity, while HEM~\cite{ai2017learning}, AEM~\cite{ai2019zero}, ZAM~\cite{ai2019zero}, and TEM~\cite{bi2020transformer} incorporate user data. Multi-scenario methods combine search and recommendation: SESRec~\cite{SESRec} uses contrastive learning, UnifiedSSR~\cite{xie2023unifiedssr} has dual-branch networks, and UniSAR~\cite{shi2024unisar} employs transformers. However, they ignores the value of consultation in enchancing search intent.

In e-commerce, \citet{zeng2025cite} uses conversation history and product knowledge to improve responses. \citet{ferreira2023rating} integrated conversational features with behavioral signals for dynamic weighting. Recently, consultation-enhanced personalized search methods have emerged, with MAPS~\cite{maps} as a representative approach that mines search motivations from consultation history to enhance current search queries. However, these methods do not take into account the value of consultations to users in personalized scenarios.

\section{Problem Formulation}


For each user \( u \in \mathcal{U} \), the corresponding chronologically stored user history \( \mathcal{H}_u = \{ \mathcal{S}_u, \mathcal{C}_u, \mathcal{D}_u \} \subseteq \mathcal{H} \) includes: (1) search session history \( \mathcal{S}_u \), (2) consultation history \( \mathcal{C}_u \), and (3) interaction history \( \mathcal{D}_u \). Specifically, \( \mathcal{S}_u = \{ s_u^{(1)} , \ldots, s_u^{(N)}  \} \subseteq \mathcal{S} \), where \( s_u^{(i)} = (q_u^{(i)}, a_{qi})\) denotes the \( i \)-th search session. $q_u^{{(i)}}$ denotes the $i$-th search query, and $a_{qi}$ denotes the $i$-th search action. Similarly, \( \mathcal{C}_u = \{ c_u^{(1)}, \ldots, c_u^{(M)} \} \subseteq \mathcal{C} \) represents \( M \) consultation sessions for user \( u \). The interaction history \( \mathcal{D}_u = \mathcal{S}_u \bigcup \{ (v_u^{(1)}, a_{v1}), \ldots, (v_u^{(K)}, a_{vK}) \} \subseteq \mathcal{D} \), which contains interacted search session history $\mathcal{S}_u$ and item \( v \in \mathcal{V} \) and corresponding interaction action categories \( a \in \mathcal{A} \).

The task of personalized search is: Given \( \mathcal{H}_u \), a new query \( q_u^{(N+1)} \), and a candidate  item list \( \mathcal{V}_u' \subseteq \mathcal{V} \), assign a ranking probability score \( p(v' | \mathcal{H}_u, q_u^{(N+1)}, \mathcal{V}_u') \) to each candidate item \( v' \in \mathcal{V}_u' \).

\section{Methodology}

In  consultation value assessment, considering that (1) value lacks explicit labels (making it difficult for model learning), and (2) consultation value is not fixed but relative (the same consultation may have different value for different searches), we therefore introduce empirical functions for assess multi-value on the data side, followed by alignment on the model side, rather than direct model learning.

\begin{figure}
    \centering
\includegraphics[width=0.48\textwidth]{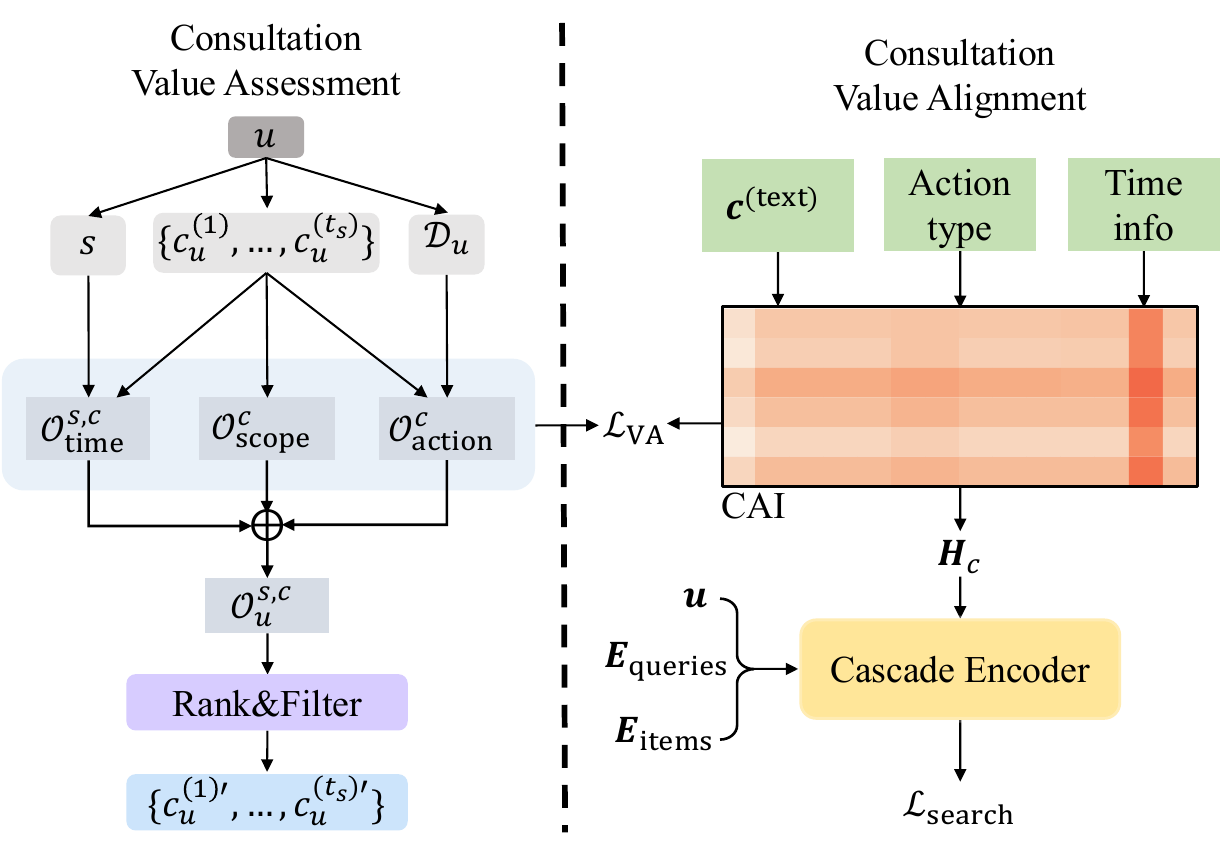}
    \caption{Overview of VAPS.}
    \label{fig:sec:method:overview}
\end{figure}

The overview of our VAPS can be found in Fig.~\ref{fig:sec:method:overview}.
We will introduce the main methods of VAPS in the next two sections: (1) Data-side Consultation Value Assessment, (2) Model-side Consultation Value Alignment.


\subsection{Consultation Value Assessment}
\label{sec:method-1}
In response to the consultation value mentioned in Sec.~\ref{sec:intro}, we first define the consultation value  as $\mathcal{O}$. 


For $\forall s \in \mathcal{S}_u$, there exist a consultation set $\mathcal{C}_u^{(t_s)} \subseteq \mathcal{C}_u$ that occurred before $s$ and an item interaction history $\mathcal{D}_u^{(t_s)} \subseteq \mathcal{D}_u$ that occurred after $s$, where $t_s$ is the timestamp of $s$. For $\forall c \in \mathcal{C}_u^{(t_s)}$, based on the target search $s$, we will present different value definitions for consultation $c$ in this section, including \textbf{time decay value}, \textbf{scenario scope value}, and \textbf{posterior action value}, all in the form of functions. Finally, the value of consultation $c$ for user $u$ and search session $s$ can be defined as $\mathcal{O}^{s,c}_u$.



\subsubsection{Time Decay Value}
\label{sec:Time Decay Value}
Inspired by forgetting curves~\cite{rubin1996one}, the \textbf{time decay value} $\mathcal{O}_{\mathrm{time}}^{s,c}$ measures the freshness value of a consultation for a user conducting a current search query. This value models the intuition that recent consultations leave a stronger impression on users and reflect their immediate needs, whereas older consultations are less relevant to their current interests. Specifically, $\mathcal{O}_{\mathrm{time}}^{s,c}$ depends on the time interval between the consultation and the current search:
$$
\mathcal{O}_{\mathrm{time}}^{s,c} = \alpha^{t_{s} - t_{c}}
$$
where $t_s$ and $t_c$ are the hour-level timestamps of $s$ and $c$, respectively. $\alpha$ is a temporal decay factor\footnote{ Based on empirical experience, for scenarios where consultations older than 30 days are deemed irrelevant, setting $\alpha = 0.99$ results in $\mathcal{O}_{\mathrm{time}}^{s,c} = 7 \times 10^{-3} \approx 0$.}.



We observe that directly using raw timestamps leads to severe sparsity issues, as each timestamp is often unique and difficult for the model to generalize across similar temporal patterns. Moreover, the time intervals between consecutive user interactions vary significantly across users, making it challenging to capture consistent temporal dependencies. Inspired by the work~\cite{li2020Time}, we discretize the time intervals into $b$ coarse-grained buckets, which helps alleviate the sparsity problem and enables the model to learn user behavior dynamics in a more robust and generalizable manner.

\subsubsection{Scenario Scope Value}
\label{sec:Scenario Scope Value}
An observation is that out-of-scope consultations (e.g., discussions about politics) provide little value to e-commerce platforms, while consultations with scenario-specific terms are more likely to accurately represent user needs. To operationalize this insight, we quantify  \textbf{scenario scope value} $\mathcal{O}_{\mathrm{scope}}^c$ with scenario-specific terms.

Terms can be directly obtained from the inverted index $\mathcal{I}$ that e-commerce search systems always maintain~\cite{catena2014inverted}. The inverted index $\mathcal{I}_{\mathrm{invert}} = \{term : [v_1,v_2,...,v_{N_t}]\}$ establishes a mapping from scenario-specific terms $term$ (e.g., ``2023 Sales Event'' and ``Folding Phone'') to their corresponding item lists, where $N_{term}$ is the total number of items linked to term $term$.

As our purpose is to eliminate out-of-scope consultations rather than to discriminate between `marginally relevant' and `highly relevant' cases,  we apply a thresholding approach to compute $\mathcal{O}_{\mathrm{scope}}^c$:
\begin{equation*}
\begin{aligned}
&\mathcal{O}_{\mathrm{scope}}^c = f_{\text{scope}}(|\mathcal{I}_c|), \quad \mathcal{I}_c \subseteq \mathcal{I}_{\mathrm{invert}} \\
&f_{\text{scope}}(x) = 
\begin{cases}
x/\lambda_{\mathrm{thresh}}, & \text{if } x < \lambda_{\mathrm{thresh}} \\
1.0, & \text{otherwise}
\end{cases}
\end{aligned}
\end{equation*}


\subsubsection{Posterior Action Value}
\label{sec:Posterior Action Value}

To distinguish between casual consultation and those that users genuinely care about, we measure the \textbf{posterior action value} $\mathcal{O}_{\mathrm{action}}^{c}$ of the target consultation $c$ through relevant consumer actions.

Here, we consider the three most typical consumer actions\footnote{For convenience, we consider only these three typical actions. Note that other actions such as `favorites' and `cart additions' are also compatible.} in e-commerce search systems: buy, click, and search, i.e., $\mathcal{A} = \{\text{search, click, buy}\}$. Here, we employ offline inverted retrieval to annotate subsequent actions associated with each consultation, with details in App.~\ref{app:details:related:actions}.

The relative importance of consumer actions varies depending on platform strategies. Based on related studies~\cite{verhallen1982scarcity, teubner2020only} and practical experience, we adhere to a common principle: Scarcer consumer actions generally hold greater value. For instance, relatively scarce purchases typically prove more valuable than relatively common clicks. The specific formula is as follows:
$$
\mathcal{O}_{\mathrm{action}}^{c} = \sum_a{\gamma_a R_{c}^{T_a}}, \quad \gamma_{a} = \frac{1}{|T_a|} \cdot \frac{1}{\sum_j\frac{1}{|T_j|}},
$$
where, $\gamma_{a}$ represents the unit scarcity value of action type $a$, $a \in \mathcal{A}$. $T_a \subseteq \mathcal{D}_u^{(t_s)}$ is the subset of interaction histories in $\mathcal{D}_u^{(t_s)}$ with action type $a$. $R_c^{T_a}$ is the action frequency of action type $a$ corresponding to the current consultation.  Details can be found in App.~\ref{app:details:bucket}.





\subsubsection{Aggregated Consultation Value}
To ensure interpretability, avoid over-penalizing low scores, and mitigate delayed feedback issue\footnote{For example, when posterior actions from recent consultations have not yet been counted, freshness value (time-decay value) is used as a compensatory measure.}~\cite{joulani2013online}, we use weighted summation. The aggregated consultation value score is computed as:
\begin{equation}
\begin{aligned}
\mathcal{O}_{u}^{s,c} = & (1-\lambda_{1})\mathcal{O}_{\mathrm{time}}^{s,c} +\\
&\lambda_{1}(\lambda_2\mathcal{O}_{\mathrm{scope}}^c + (1-\lambda_2)\mathcal{O}_{\mathrm{action}}^{c})
\end{aligned}
\label{eq:method:aggregated}
\end{equation}
where $\lambda_1$ and $\lambda_2$ are hyper-parameters controlling the weight (analysis in Sec.~\ref{sec:exp:hyperparameter}). 

Considering the maximum input length $L_\text{seq}$ for model~\cite{maps,shi2024unisar}, we can rank and filter the consultation history for each search session $s$ of the user to get rid of useless consultation:
$$
\mathcal{C}'_u = \text{Rank\&Filter}_{\mathcal{O}_{u}^{s,c}}(\mathcal{C}_u,L_\text{seq},s,u).
$$




\subsection{Model-side Consultation Value Alignment}
\label{sec:method-2}

Previous works only consider the similarity between consultation history and the current query, without integrating interactions with user consumption actions on the model side. Based on the data-side consultation value assessment in Sec.~\ref{sec:method-1}, we expect the model to understand consultation needs not only by measuring the semantic similarity between consultations and the current query but also by modeling the interactions between consultations and user consumption actions to assess whether and to what extent consultations are valued and recognized by users. Therefore, to enable the model to understand the correlation between user consumption actions and consultation history, we introduce \textbf{Consultation-Action Interaction (CAI)}, aiming to help the model learn whether (and how strongly) consultations are supported by corresponding user consumption actions, thereby inferring the degree of user recognition for consultations.

\subsubsection{Embedding Construction}

First, based on the same dimensions $d$, we construct a token embedding layer $\text{TokenEmb}()$ for the text vocabulary to model textual semantics, along with item embedding and user embedding layers ($\text{ItemEmb}()$ and $\text{UserEmb}()$) to store collaborative information of items and users, respectively. 

For user $u$, the user embedding $\bm u = \text{UserEmb}(u)$. Given the $u$'s ranked and filtered consultation history $\mathcal{C}'_u=\{c_{u}^{(1)}, ...,c_{u}^{(M')}\}$, we can obtain the text embedding for consultations through MoAE (Mixture of Attention Expert) text encoder~\cite{maps}:
\begin{equation*}
    \bm{c}^\text{(text)}_{j} = \text{MoAE}(\text{TokenEmb}(tok_{j,1},\dots,tok_{j,N_{tok}})),
\end{equation*}
where $\text{tok}_{j,k}$ is the $k$-th token of the text $c_u^{(j)}$, $j = 1, 2, \dots, M'$. We obtain the user consultation text sequence:
$[c^\text{(text)}_j; \dots; c^\text{(text)}_M]$. Similarly, we can obtain their corresponding text embeddings $q_j^{\text{(text)}}, j=1,2,\dots,N$ and $v_j^{\text{(text)}}, j=1,2,\dots,K$.

\subsubsection{Consultation-Action Interaction}

Inspired by~\citet{lin2022cat}, we propose Consultation-Action Interaction (CAI), which employs cross-attention between consumer actions 
(as keys/values) and consultations (as queries). The purpose is to identify posterior support in consumer actions for consultations and capture the underlying relationships between consultations and actions.

Here, we introduce (1) discrete time interval embeddings $\text{TimeEmb}()$ to align with the temporal awareness in Sec.~\ref{sec:Time Decay Value}, and (2) action-type embeddings $\text{ActionEmb}()$ to distinguish between different action types as mentinoed in Sec.~\ref{sec:Posterior Action Value}.
For $a_i$ in $\mathcal{D}_u$, the embedding is:
$$
\bm a_i = \begin{cases} \text{ActionEmb}(a_i)  + \bm{v}_{a_i} & \text{if }a \text{ in \{buy, click\}}\\  \text{ActionEmb}(a_i)  + \bm{q}_{a_i} & \text{if }a \text{ in \{search\}} \end{cases}.
$$
We further incorporate temporal information to form the attention inputs:
\begin{equation*}
\begin{aligned}
&\bm{e}^{\text{(attn-q)}}_{i}= \bm c^\text{(text)}_{i} + \text{TimeEmb}(c_{i}), \\
&\bm{e}^{\text{(attn-k)}}_{j}=\bm{e}^{\text{(attn-v)}}_{j} = \bm a_i + \text{TimeEmb}(a_j).
\end{aligned}
\end{equation*}

Subsequently, the cross-attention mechanism derives posterior support from actions for $c_i$, while preserving the original text semantics through skip connections with hyperparameter $\lambda_3$:
$$
\bm{h}_{c_i} = \bm c^\text{(text)}_{i} + \lambda_3\text{Attn}(\bm{e}^{\text{(attn-q)}}_i, \bm{E}^{\text{(attn-k)}}, \bm{E}^{\text{(attn-v)}}),
$$
where $\bm E^{\text{(attn-k)}} = [\bm{e}^{\text{(attn-k)}}_{1}; ... ;\bm{e}^{\text{(attn-k)}}_{K}],\ \bm E^{\text{(attn-v)}} = [\bm{e}^{\text{(attn-v)}}_{1}; ... ;\bm{e}^{\text{(attn-v)}}_{K}]$.

\subsubsection{Value-Assessment Alignment}
To ensure alignment with value assessment in Sec.~\ref{sec:method-1}, besides ranking and filtering consultations with $\mathcal{O}_{u}^{s,c}$, we propose a value-assessment alignment loss $\mathcal{L}_{\text{VA}}$ to supervise the cross-attention scores of CAI.

Specifically, we obtain consultation-action pairs for the consultation-action 
 mapping in  Sec.~\ref{sec:Posterior Action Value}. For each pair $(c, a)$, we compute the weight-projected embeddings embeddings: $\bm{e}^{\text{(attn-q)}'} = \bm{e}^{\text{(attn-q)}} \bm{W}^\text{(attn-q)}$, $\bm{e}^{\text{(attn-k)}'} = \bm{e}^{\text{(attn-k)}} \bm{W}^\text{(attn-k)}$.
Considering that the cross attention scores are obtained through dot product and softmax, and inspired by MAPS's alignment, we employ softmax-based contrastive learning to supervise the attention:
 \begin{equation*}
 \begin{aligned}
     &\text{sim}(c,a) =  \bm{e}^{\text{(attn-q)}'}\cdot\bm{e}^{\text{(attn-k)}'} \\
      & \mathcal{L}_{\text{VA}} = -\sum_{(c,a)}\text{log}\frac{\text{exp}(\text{sim}(c,a)/\tau_\text{1})}{\sum_{a^-\in \mathcal{D}_u\backslash a}\text{exp}(\text{sim}(c,a^-)/\tau_\text{1})}
 \end{aligned}
 \end{equation*}
where $\tau_1$ is temperature parameter used to control the sharpness of the softmax distribution~\citep{hinton2015distilling}.
\subsubsection{Personalized Search Learning}

Through the cascaded bidirectional attention encoder~\cite{maps}, we obtain the final query embedding.
$$
\bm e^\text{(final)}_q = \text{Cascaded-Encoder}
(\bm H_c, \bm E_{\text{items}},\bm E_{\text{queries}},\bm u),
$$
where $\bm{H}_{c} = [\bm h_{c_1};\dots;\bm h_{c_M}]$, $\bm E_{\text{items}} = [\bm v_1;\dots;\bm v_K]$, $\bm E_{\text{queries}} =[\bm q_1;\dots;\bm q_N]$. For inference, candidate item $\bm v'$  can be ranked based on similarity-derived probability scores:
$$
p(v'|\mathcal{H}_u, s_u^{(N+1)}, \mathcal{V}_{u}') = \text{sim}(\bm e^\text{(final)}_q, \bm v').
$$

For optimization, following previous works~\citep{bi2020transformer,ai2017learning,shi2024unisar}, the objective is to increase the similarity scores of ground-truth given user history. The personalized alignment loss $\mathcal{L}_\text{search}$ can be formulated as:
\begin{equation*}
   \mathcal{L}_\text{search} = \sum_{(u, v, q)}\text{log}\frac{\text{exp}(\text{sim}(\bm e^\text{(final)}_q, \bm v)/\tau_\text{2})}{\sum_{v'\in\mathcal{V}_{u}'}\text{exp}(\text{sim}(\bm e^\text{(final)}_q, \bm v')/\tau_\text{2})}.
\end{equation*}
Following existing works~\cite{ai2019explainable,shi2024unisar}, we employ negative sampling~\citep{le2014distributed}. The overall loss $\mathcal{L}_{\text{final}}$ is:
\begin{equation*}
\mathcal{L}_{\text{final}} = \mathcal{L}_\text{search} + \lambda_3 \mathcal{L}_\text{VA} + \lambda_4 ||\Theta||_2,
\end{equation*}
where $\lambda_3$ and $\lambda_4$ is a hyper-parameters, $\Theta$ is model parameters.

\section{Experiment}
 We answer the following research questions in this section: 
 \textbf{RQ1:} How does VAPS rank compared to existing baselines?
 \textbf{RQ2:} How effective is VAPS in retrieval?
 \textbf{RQ3:} How does VAPS compare to multi-scenario methods?
  \textbf{RQ4:} How effective are VAPS's individual modules?
   \textbf{RQ5:} How reliably does VAPS assess consultation value?
   \textbf{RQ6:} How do hyper-parameters influence consultation value assessment in VAPS?
  \textbf{RQ7: } What is the time complexity of VAPS?

\subsection{Experiment Settings}
\subsubsection{Datasets} 

To validate VAPS's effectiveness, experiments are conducted on two datasets. \textbf{Commercial dataset} is a real user interaction dataset from an e-commerce platform with AI consulting services~\cite{maps}. We follow the original setup by using the last two days of data for validation and testing.
\textbf{Amazon dataset} is derived from the Amazon Reviews dataset~\citep{ni2019justifying}, subsequently enhanced by PersonalWAB~\citep{cai2024personalwab} and MAPS~\cite{maps}, containing user profiles as well as interaction behaviors such as searches and consultations. We adopt the dataset version used in MAPS and strictly follow its data processing. The statistics of these datasets are shown in Tab.~\ref{tab:exp:statistics}

\begin{table}[!tbp]
\centering
\setlength{\tabcolsep}{3.4pt}\fontsize{9}{7}\selectfont
\begin{tabular}{lcccc}
\toprule
\textbf{Dataset}         & \textbf{\#Users}   & \textbf{\#Items}   & \textbf{\#Inters}      \\ \midrule
\textbf{Commercial}    & 2096           & 2691  & 24662 / 18774               \\
\textbf{Amazon}   & 967           & 35772 & 7263 / 40567                 \\
\bottomrule
\end{tabular}
\caption{Statistics of the 2 pre-processed datasets. In ``\#Inters'', searches are shown on the left of `/', and consultations are shown on the right.}
\label{tab:exp:statistics}
\end{table}

\subsubsection{Baselines}
For \textbf{ranking} evaluation, we adopt these personalized search baselines: \textbf{ZAM}~\citep{ai2019zero}, \textbf{HEM}~\citep{ai2017learning}, \textbf{AEM}~\citep{ai2019zero}, \textbf{QEM}~\citep{ai2019zero}, \textbf{TEM}~\citep{bi2020transformer},\textbf{CoPPS}~\citep{CoPPS}, and \text{MAPS}~\citep{maps}. For \textbf{retrieval} performance, we additionally introduce traditional, dense, and conversational retrieval methods: \textbf{BM25}~\citep{bm25}, \textbf{BGE-M3}~\citep{chen2024bge}, and \textbf{CHIQ}~\citep{mo2024chiq}. Furthermore, we also consider include multi-scenario methods, including \textbf{SESRec}~\citep{SESRec}, \textbf{UnifiedSSR}~\citep{xie2023unifiedssr}, and \textbf{UniSAR}~\citep{shi2024unisar} . For more model settings and implementation details, see App.~\ref{app:exp:baselines}.

\subsubsection{Metrics and Implementation details}

Following~\cite{shi2024unisar,zhang2024qagcf,zhang2024modeling}, we adopt \textbf{Hit Ratio} (HR@$k$) and \textbf{Normalized Discounted Cumulative Gain} (NDCG@$k$ or N@$k$) for ranking, and \textbf{Mean Reciprocal Rank} (MRR@$k$) for retrieval. Following~\cite{maps,shi2025unified,zhang2024saqrec}, each ground-truth item is paired with 99 negatives, evaluating HR/NDCG at $\{5,10,20,50\}$. For retrieval, all items are candidates with MRR reported at $\{10,20,50\}$. Details appear in App.~\ref{app:exp:implement}.



\begin{table*}[t]
\centering
\small
\begin{tabular}{l|cccccccc}
\toprule
Model         & HR@5            & HR@10           & HR@20           & HR@50           & NDCG@5          & NDCG@10         & NDCG@20         & NDCG@50         \\ \midrule
\multicolumn{9}{c}{Commercial}                                                                                                                                \\ \midrule
ZAM           & 0.3680          & 0.5247          & 0.6810          & 0.8203          & 0.2491          & 0.2988          & 0.3378          & 0.3659          \\
HEM           & 0.3487          & 0.4911          & 0.6375          & 0.8036          & 0.2359          & 0.2811          & 0.3183          & 0.3522          \\
AEM           & 0.3892          & 0.5372          & 0.6721          & 0.8255          & 0.2648          & 0.3127          & 0.3474          & 0.3790          \\
QEM           & 0.3991          & 0.5468          & 0.6732          & 0.8438          & 0.2675          & 0.3148          & 0.3461          & 0.3803          \\
TEM           & 0.4062          & 0.5681          & 0.7199          & 0.8742          & 0.2869          & 0.3405          & 0.3761          & 0.4053          \\
CoPPS         & 0.4057          & 0.5632          & 0.7178          & 0.8656          & 0.2829          & 0.3345          & 0.3735          & 0.4033          \\
MAPS          & 0.5276          & 0.7064          & 0.8321          & 0.9323          & 0.3762          & 0.4360          & 0.4639          & 0.4871          \\
\textbf{VAPS} & \textbf{0.5565}$^\dagger$   & \textbf{0.7145}$^\dagger$ & \textbf{0.8398} & \textbf{0.9422}$^\dagger$ & \textbf{0.3884}$^\dagger$ & \textbf{0.4424}$^\dagger$ & \textbf{0.4726}$^\dagger$ & \textbf{0.4931}$^\dagger$ \\ \midrule
\multicolumn{9}{c}{Amazon}                                                                                                                                    \\ \midrule
ZAM           & 0.3100          & 0.4487          & 0.5433          & 0.7302          & 0.1826          & 0.2115          & 0.2498          & 0.2782          \\
HEM           & 0.2736          & 0.4192          & 0.5412          & 0.7458          & 0.1984          & 0.2173          & 0.2595          & 0.2975          \\
AEM           & 0.3184          & 0.4559          & 0.5366          & 0.7247          & 0.1861          & 0.2128          & 0.2470          & 0.2769          \\
QEM           & 0.2832          & 0.3879          & 0.5283          & 0.7664          & 0.1900          & 0.2122          & 0.2273          & 0.2918          \\
TEM           & 0.4028          & 0.4813          & 0.7201          & 0.8051          & 0.2965          & 0.3123          & 0.3416          & 0.3540          \\
CoPPS         & 0.3871          & 0.4862          & 0.7289          & 0.8013          & 0.2784          & 0.3299          & 0.3435          & 0.3696          \\
MAPS          & 0.6062          & 0.7835          & 0.8990          & 0.9702          & 0.4237          & 0.4717          & 0.5001          & 0.5189          \\
\textbf{VAPS} & \textbf{0.6418}$^\dagger$ & \textbf{0.8019}$^\dagger$ & \textbf{0.9101}$^\dagger$ & \textbf{0.9748} & \textbf{0.4903}$^\dagger$ & \textbf{0.5213}$^\dagger$ & \textbf{0.5556}$^\dagger$ & \textbf{0.5665}$^\dagger$ \\ \bottomrule
\end{tabular}
\caption{Search ranking performance compared with personalized search baselines. The best results are shown in bold. `$\dagger$' indicates the model significantly outperforms all baseline models with paired t-tests at $p < 0.05$ level.}
\label{tab:exp:ranking}
\end{table*}

\begin{table}[t]
\centering
\small
\begin{tabular}{@{}l|cccc@{}}
\toprule
Method & MRR@5  & MRR@10 & MRR@20 & MRR@50 \\ \midrule
ZAM    & 0.2211 & 0.2539 & 0.2628 & 0.2702 \\
HEM    & 0.2251 & 0.2524 & 0.2739 & 0.2872 \\
AEM    & 0.2132 & 0.2424 & 0.2630 & 0.2705 \\
QEM    & 0.2264 & 0.2540 & 0.2655 & 0.2808 \\
TEM    & 0.2597 & 0.2729 & 0.3089 & 0.3239 \\
CoPPS  & 0.2517 & 0.2806 & 0.3178 & 0.3371 \\
BM25   & 0.2780 & 0.2870 & 0.2917 & 0.2997 \\
BGE-M3 & 0.3408 & 0.3540 & 0.3614 & 0.3651 \\
CHIQ   & 0.3526 & 0.3691 & 0.3883 & 0.4047 \\ 
MAPS   & 0.4119 & 0.4324 & 0.4397 & 0.4523 \\ \midrule
\textbf{VAPS} & \textbf{0.4559}$^\dagger$            & \textbf{0.4749}$^\dagger$            & \textbf{0.4815}$^\dagger$            & \textbf{0.4944}$^\dagger$            \\ \bottomrule
\end{tabular}
\caption{Retrieval performance on the Amazon dataset.}
\label{tab:exp:retrieval}
\end{table}

\begin{table}[t]
\centering
\small
\begin{tabular}{l|cccc}
\toprule
Method   & HR@10           & HR@20           & N@10         & N@20         \\ \midrule
SESRec       & 0.5601          & 0.7189          & 0.3464          & 0.3788          \\
UnifiedSSR     & 0.5709          & 0.7082          & 0.3598          & 0.3793          \\
UniSAR        & 0.5837          & 0.7295          & 0.3605          & 0.3894         \\
MAPS               & 0.7064 & 0.8321 & 0.4360 & 0.4639 \\\midrule
\textbf{VAPS}                & \textbf{0.7145}$^\dagger$ & \textbf{0.8398} & \textbf{0.4424}$^\dagger$ & \textbf{0.4726}$^\dagger$ \\
\bottomrule
\end{tabular}
\caption{Search ranking performance compared with multi-scenario baselines on the Commercial dataset.}
\label{tab:exp:cross:comp}
\end{table}

\begin{table}[t]
\centering
\setlength{\tabcolsep}{2.5pt}
\small
\begin{tabular}{l|ccccc}
\toprule
Ablation  & HR@5 & HR@10           & HR@20           & N@10         & N@20         \\ \midrule
VAPS      & \textbf{0.6418}         & \textbf{0.8019} & \textbf{0.9101} & \textbf{0.5213} & \textbf{0.5556} \\ \midrule
\ w/o $\mathcal{O}_{\mathrm{time}}^c$  & 0.6133         & 0.7784          & 0.9011          & 0.4783          & 0.5106          \\
\ w/o $\mathcal{O}_{\mathrm{scope}}^c$  & 0.6189         & 0.7853          & 0.9036          & 0.4897          & 0.5202          \\
\ w/o $\mathcal{O}_{\mathrm{action}}^c$  & 0.6128        & 0.7792          & 0.9010          & 0.4795          & 0.5135          \\ \midrule
\ w/o $\mathcal{L}_\text{VA}$ & 0.6306  & 0.7909          & 0.9051          & 0.5062          & 0.5419          \\
\ w/o $\text{CAI}$  & 0.6334  & 0.7867         & 0.9049           & 0.4928   & 0.5258     \\
\bottomrule
\end{tabular}
\caption{Ablation study of VAPS on Amazon.}
\label{tab:exp:ablation}
\end{table}





\subsection{Overall Performance}  
 In a retrieval-then-ranking e-commerce system, the primary objective of personalized product search is personalized ranking performance. 
 To answer \textbf{RQ1}, \textbf{RQ2}, and \textbf{RQ3}, we first evaluate the ranking performance in Tab.~\ref{tab:exp:ranking} and Tab.~\ref{tab:exp:cross:comp}, followed by a comparison of the retrieval performance in Tab.~\ref{tab:exp:retrieval}. 

Regarding ranking, VAPS outperforms all other personalized product search methods and search-integrated multi-scenario approaches. The improvements are significant across most metrics, particularly on the Amazon dataset. We also observed relatively smaller gains (approximately 2\%) on Commercial. We attribute it to the inherently limited quantity of ground-truth items on Commercial and data saturation effect.
Concerning retrieval, VAPS surpasses all personalized product search methods and traditional, dense, and conversational retrieval approaches.
This fully shows VAPS' effectiveness and superiority in ranking and retrieval tasks, highlighting its ability to boost search performance on e-commerce platforms.


\subsection{Ablation Study} 
\label{sec:exp:ablation}
In this section, we discuss the specific roles of each module in VAPS to answer \textbf{RQ4}. 
As shown in Table~\ref{tab:exp:ablation}, removing $\mathcal{O}_{\text{time}}^c$ from the aggregated consultation value yields the most significant performance drop, followed by removing $\mathcal{O}_{\text{action}}^c$. $\mathcal{O}_{\text{time}}^c$ directly measures the temporal difference between consultations and the current search. Through a decay function, it effectively captures the user's evolving search-consultation interests over time. When $\mathcal{O}_{\text{time}}^c$ is removed, the aggregated consultation value loses temporal context, failing to distinguish the similarity of old versus new consultations to the current search. This results in personalized search returning outdated or irrelevant content. Conversely, removing $\mathcal{O}_{\text{action}}^c$ causes the aggregated consultation value to overlook validation from actual user actions in later interations. Relying solely on semantic matching of search queries and temporal information, the filtered consultations lack real-world posterior support, leading personalized search astray from genuine user needs.



\subsection{Consultation Value Distribution Analysis}
To answer \textbf{RQ5}, we present the assessed values in a distribution format for both datasets.

\begin{figure}[!t]
    \centering
    \begin{subfigure}[h]{0.23\textwidth}
        \includegraphics[width=\textwidth]{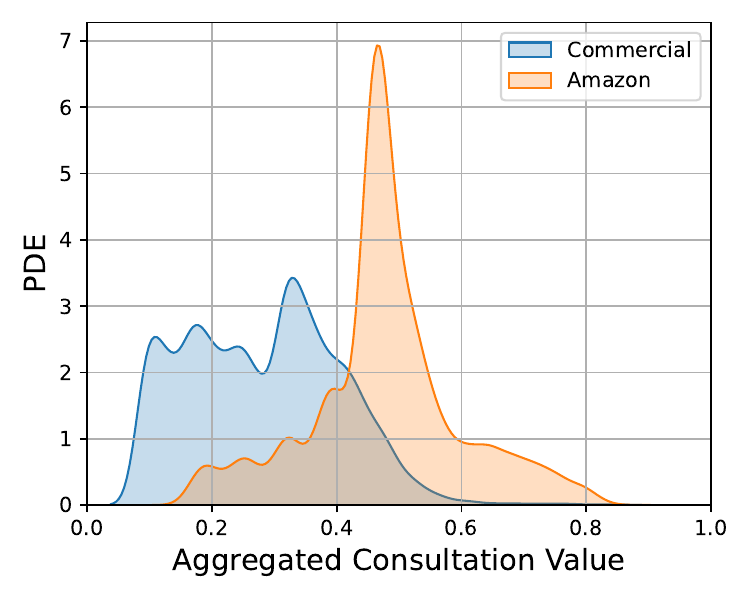}
        \caption{$\mathcal{O}_{u}^{s,c}$ on the Amazon and Commercial datasets.}
        \label{fig:exp:dist:agg}
    \end{subfigure}
    \hfill
    \begin{subfigure}[h]{0.23\textwidth}
        \includegraphics[width=\textwidth]{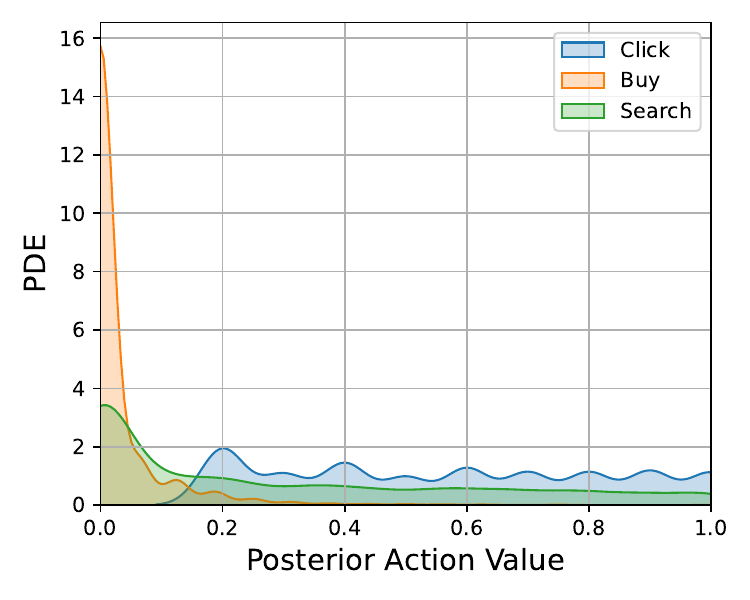}
        \caption{$\mathcal{O}_{\mathrm{action}}^{c}$ of each $a$ on Commercial.}
        \label{fig:exp:dist:action}
    \end{subfigure}
    \caption{Value distribution visualization.}
    \label{fig:exp:dist:ds}
\end{figure}

Fig.~\ref{fig:exp:dist:agg} reveals that the commercial dataset exhibits lower normalized scores overall, attributable to its sparser user interactions, consistent with the fact that its primary product categories are non-daily necessities. In Fig.~\ref{fig:exp:dist:action}, we find the value score distribution of clicks is very flat, reflecting their high prevalence as a user action. In contrast, searches are relatively concentrated near 0, while purchases exhibit the sharpest distribution, reflecting increasing scarcity. As the ``action cost'' progressively increases, action become increasingly scarce and consequently more valuable (indicating stronger user intent).

\subsection{Case Study}
In this section, we analyse discrete cases for consultation value scores in both commercial and amazon datasets.

\begin{figure}[!t]
    \centering
    \includegraphics[width=1\linewidth]{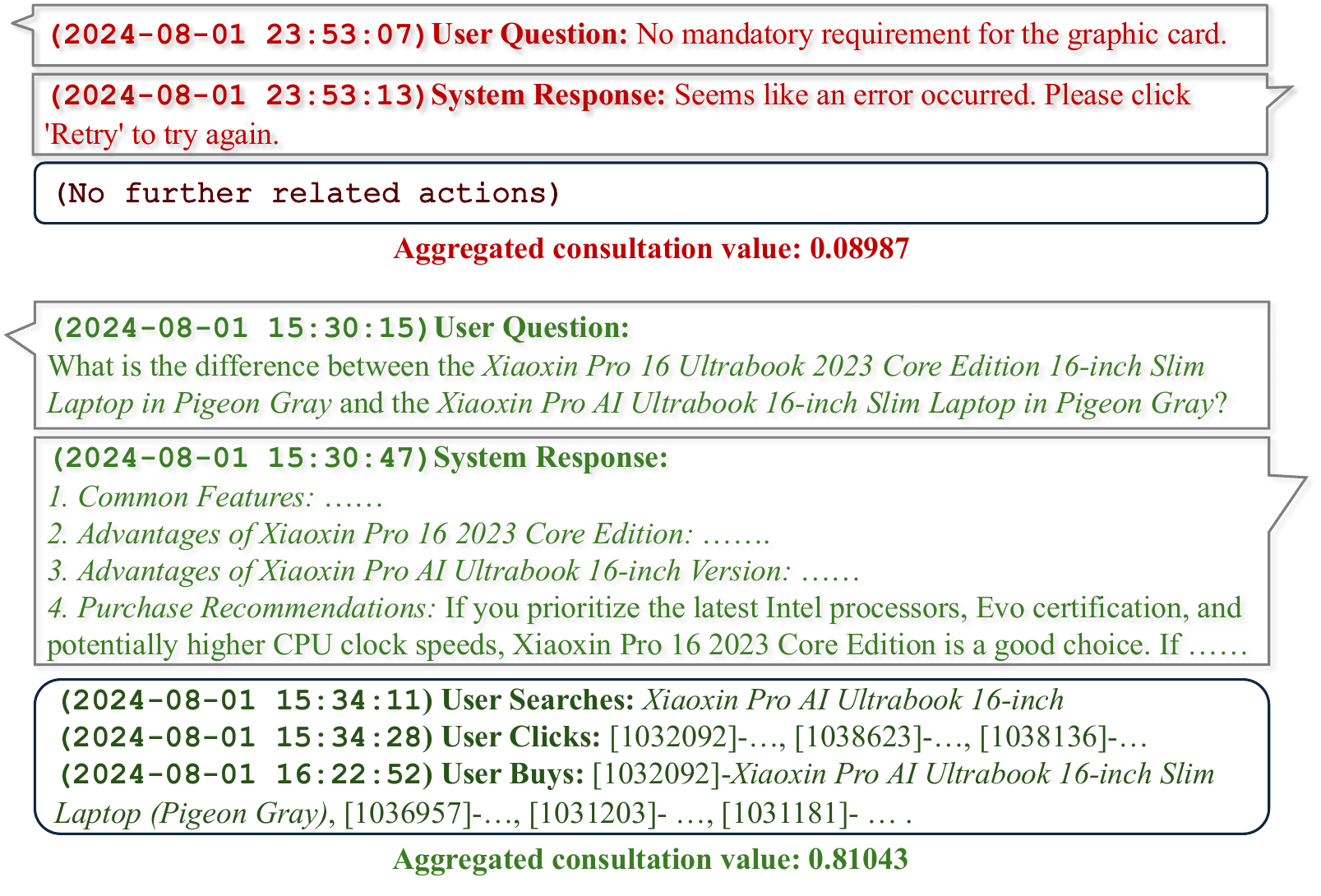}
    \caption{Consultation value case study with actions on Commercial.}
    \label{fig:exp:case:commercial}
\end{figure}

As shown in Fig.~\ref{fig:exp:case:commercial}, the consultation in the first case provide no useful information about the graphic card requirement, reflected by the low value of 0.08987. In contrast, the second case involved a user query about the differences between two laptops. The system give a comprehensive response covering common features, advantages of each model, and purchase recommendations. Moreover, the consultation is also in-scope and timely, verified by subsequent user actions such as related searches and purchases, resulting in a high-valued consultation (0.81043).

\subsection{Hyperparameter Analysis}
\label{sec:exp:hyperparameter}
In this section, we investigate the impact of changing hyper-parameters $\lambda_1$ and $\lambda_2$ to answer \textbf{RQ6}. According to Eq.~\eqref{eq:method:aggregated}, 
the values of $\lambda_1$ and $\lambda_2$ should be within the range of [0, 1]. We performed parameter tuning for $\lambda_1$ and $\lambda_2$ separately. By freezing one parameter at a time, we test multiple weight values between 0 and 1 for the other parameter. The optimal $\lambda_1$ is found to be 0.5, which highlights the necessity of analyzing the temporal impact of consultations independently. The optimal $\lambda_2$ is 0.3, indicating that posterior actions are more critical than scenario scopes for the aggregated consultation value and have a greater influence on user behavior. Meanwhile, we assume that the scope value (based on inverted indices) might be overly restrictive, and more effective alternatives should be explored in future work.

These conclusions are consistent with those in Sec.~\ref{sec:exp:ablation}.
\begin{figure}[!t]
    \centering
    \begin{subfigure}[h]{0.23\textwidth}
        \includegraphics[width=\textwidth]{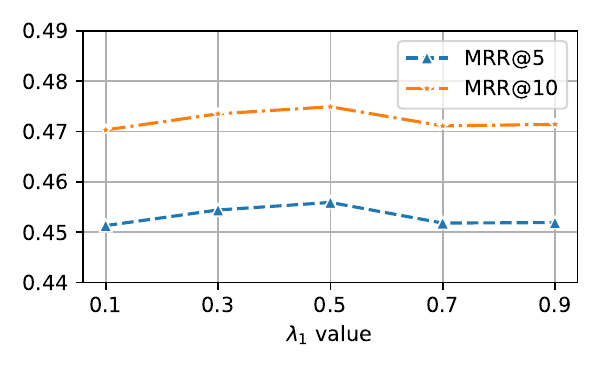}
        \caption{$\lambda_1$}
        \label{fig:exp:hyper:lambda:1}
    \end{subfigure}
    \hfill
    \begin{subfigure}[h]{0.23\textwidth}
        \includegraphics[width=\textwidth]{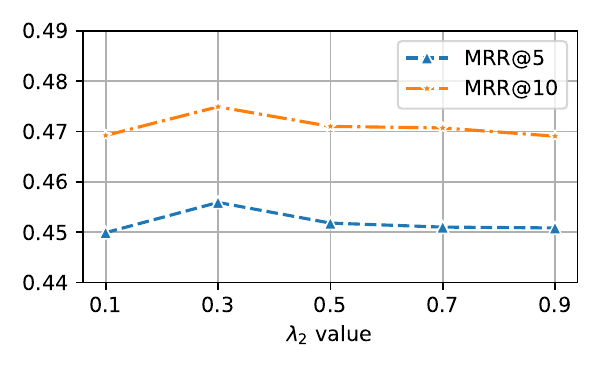}
        \caption{$\lambda_2$}
        \label{fig:exp:hyper:lambda:2}
    \end{subfigure}
    \caption{Impact of different hyperparameters on Amazon. Default setting is $\lambda_1=0.5,\lambda_2=0.3$.}
    \label{fig:exp:hyper:lambda}
\end{figure}

\subsection{Complexity Analysis}
To address \textbf{RQ7}, we conduct complexity analysis in this section. VAPS has an overall time complexity of $O(N'^2d)$ ($N'=\text{max}(M,N,K)$; see App.~\ref{app:exp:complexity} for details), matching state-of-the-art methods like UniSAR and MAPS. Notably, VAPS outperforms these methods under the same complexity constraints. While optimizing computational efficiency for large-scale platforms matters, we defer this to future work.

\section{Conclusion}

We propose the VAPS method to enhance e-commerce personalized search by leveraging user consultations. Existing semantic-only methods fail to capture consultation value. To address this gap, we propose a consultation value assessment framework comprising three dimensions: Scenario Scope, Posterior Action, and Time Decay Value. Based on this framework, we develop VAPS, a value-aware personalized search model that incorporates high-value consultations through a consultation–user action interaction module and an objective function explicitly aligning consultations with user actions. 

\section{Limitations}

The VAPS framework advances personalized search through its novel consultation value assessment and alignment mechanisms, though there remains some limitations and room for improvement in certain aspects to further enhance its research contributions. First, while its time complexity aligns with state-of-the-art methods, further optimizing the computation for ultra-large-scale environments could enhance real-time deployment flexibility. Regarding hyperparameter tuning, while the optimal values of hyper-parameters are validated in experimental settings, developing automated tuning strategies could reduce dependency on manual adjustments across diverse business scenarios. Furthermore, the model's reliance on historical interaction data raises considerations for cold-start scenarios. Integrating content-based features or transfer learning could mitigate this challenge. Lastly, while the current design focuses on interaction data, enriching the framework with external knowledge graphs or large language model insights could enhance its ability to parse complex consultative intents in specialized domains. These observations highlight opportunities for future research to further strengthen VAPS's scalability, adaptability, and semantic modeling capabilities.

\bibliography{custom}

\appendix


\section{Appendix: Methodology Details}
\subsection{Appendix: Details of Related Actions For Consultations}
\label{app:details:related:actions}
Inspired by \citet{maps}, to build an \textbf{offline inverted retrieval} for recording the relevant consumer actions corresponding to each consultation, we set the following conditions and rules:  
1. For each consumer action \( a \in \mathcal{A} \) of a user, we search for related consultations \( c \in \mathcal{C} \) within a certain time period before the action occurs (time ranges considered: 3, 5, 7, 14 days; due to the sparsity of the dataset, we set the time range to 14 days), and construct an action-related consultation table \( x = \{ a : \mathcal{C}_{\text{related}} \} \). Specifically, if the text information \( {TI} \) corresponding to the consumer action \( a \) (e.g., search query; item name and text features for clicks/purchases) appears completely at least once in consultation \( c \), or more than half of the item contents in $TI$ appears in \( c \) as text, or more than half of the query terms in \( {TI} \) appear in \( c \) as text, then consultation \( c \) is added to \( \mathcal{C}_{\text{related}} \).  
2. Reverse each key-value pair in the action-related consultation table \( x \) to obtain a consultation-related action inverted table \( y = \{ c : \mathcal{A}_{\text{related}} \} \), which constitutes the final content of the offline inverted retrieval.

\subsection{Appendix: Details of the Usage of Bucketization and $R_c^{T_a}$ for Posterior Action Values}
\label{app:details:bucket}
To enhance the robustness of subsequent model training, the calculated consultation value has been normalized to [0, 1]. Consequently, each component of the consultation value should also be normalized to [0, 1]. While the normalization of time decay and scenario scope value is relatively straightforward, typically using direct exponential or reciprocal-based methods, the normalization of posterior action value needs additional consideration.

In e-commerce platforms, consumer action data (e.g., user clicks, purchases) often exhibit long-tailed distributions, where a few extreme values \cite{chen2023long, jin2023brush} (e.g., a hacker artificially generating over 10k clicks in short intervals through network attacks) can significantly distort normalization methods sensitive to boundary conditions, such as Min-Max and Z-Score \cite{henderi2021comparison}. Bucketization, however, discretizes continuous values into fixed intervals, effectively mitigating the impact of outliers~\cite{jahrer2010combining, cao2011sabre}.

Therefore, we introduce bucketization in posterior action value part of VAPS to convert raw action counts into relative frequency values \( R_c^{T_a}\)), which includes:

1. \textbf{Construct Action Frequency Table }\( \text{freq}_\mathcal{A} \):  
   Using the consultation-related action inverted table \( y = \{c : \mathcal{A}_{\text{related}}\} \) from offline inverted retrieval, we count the occurrences of each action type for every consultation \( c \), aggregating them into a consultation-action frequency table \( \text{freq}_\mathcal{A} \).

2. \textbf{Generate Eleven Equal Quantiles}:  
   For each action type in \( \text{freq}_\mathcal{A} \), compute the eleven equal quantiles (i.e., dividing the data into 11 groups with equal probability mass). These quantiles define 11 buckets (No.0 to No.10) for each action type.

3. \textbf{Map Frequencies to Relative Values}:  
   For a given consultation \( c \) and its corresponding posterior action frequency \( T_a \), determine the bucket to which \( T_a \) belongs. The relative frequency value \( R_c^{T_a} \) is then calculated as the bucket index divided by 10 (e.g., bucket No.5 maps to \( R_c^{T_a} = 0.5 \)).

\section{Appendix:  Experiment Details}


\subsection{Baseline Details}
\label{app:exp:baselines}
We initiate our comparisons by evaluating our method against various baselines for ranking and retrieval tasks. For personalized search models, we consider the following:
\begin{itemize}
\item \textbf{AEM}~\cite{ai2019zero}, an attention-based personalized model that merges the user's historical interaction items with the current query, enabling a more context-aware search experience.
\item \textbf{QEM}~\cite{ai2019zero}, which solely focuses on the matching scores between items and queries, providing a more straightforward but query-centric ranking approach.
\item \textbf{HEM}~\cite{ai2017learning}, a latent vector-based personalized model that captures user preferences through hidden representations.
\item \textbf{ZAM}~\cite{ai2019zero}, an enhanced version of AEM that appends a zero vector to the item list, aiming to improve the model's performance in certain scenarios.
\item \textbf{TEM}~\cite{bi2020transformer}, which replaces the attention layer of AEM with a transformer encoder, leveraging the powerful sequential modeling capabilities of transformers.
\item \textbf{CoPPS}~\cite{CoPPS}, a model that harnesses contrastive learning techniques to enhance personalized search performance.
\end{itemize}
In addition, we benchmark our method against \textbf{MAPS}~\cite{maps}, the pioneering model that utilizes consultation information for personalized search.
Next, we compare our approach with multi-scenario methods that integrate search and recommendation interactions:
\begin{itemize}
\item \textbf{SESRec}~\cite{SESRec} employs contrastive learning to learn disentangled search representations, facilitating more effective recommendations.
\item \textbf{UnifiedSSR}~\cite{xie2023unifiedssr} jointly models user behavior history across search and recommendation scenarios, capturing the intricate relationships between the two.
\item \textbf{UniSAR}~\cite{shi2024unisar} effectively models diverse fine-grained behavior transitions using two distinct transformers and implements a cross-attention mechanism for enhanced interaction modeling.
\end{itemize}
For the retrieval task, we contrast VAPS with traditional, deep learning-based, and conversational-based baselines:
\begin{itemize}
\item \textbf{BM25} relies on word frequency to identify and rank relevant retrieval candidates, a classic approach in information retrieval.
\item \textbf{BGE-M3} incorporates concepts such as embedding to boost the performance of retrieval tasks, leveraging deep learning techniques.
\item \textbf{CHIQ} endeavors to integrate world knowledge from large language models (LLMs) into the search process, aiming to enhance retrieval effectiveness.
\end{itemize}

\subsection{Implementation Details}
\label{app:exp:implement}

All hyperparameters of the baseline are searched according to the settings in the original paper. 
Following related work~\citep{shi2024unisar}, we set $d$ to 64, and the maximum length $L_\text{seq}$ of the user history sequence to 30. Fully following~\citet{maps}, we filter out users with fewer than 5 interactions, use `tanh' as the activation function, set the number of layers in the transformer encoder to 1, batch size to 72, and the number of negative samples for each positive sample for $\mathcal{L}_\text{search}$ to 10. For $\mathcal{L}_\text{VA}$, the random sampled negative is adopted, with the corresponding batch size searched among \{64,128,256\}. $\lambda_{thresh}$ is set to 4.
$\tau_1$, $\tau_2$, $\lambda_1$, $\lambda_2$, $\lambda_3$ and $\lambda_4$ is tuned in \{0: 0.1: 1\}. $\alpha$ is tuned in \{0.98,0.99,0.999,0.9999\}.
We train all models for 100 epochs with early stopping (5 epochs) to avoid overfitting and optimize using Adam~\cite{kingma2014adam}.
The learning rate is adjusted among \{1e-3, 5e-4, 1e-4, 5e-5, 1e-5\}.
All experiments were completed on an A800 (80GB) GPU. 

\subsection{Appendix: Complexity Analysis Details}
\label{app:exp:complexity}
In this section, we give detailed complexity analysis of VAPS.
We analyse that time complextity of VAPS includes: (1) construction of the consultation-related action inverted table  \( y = \{c : \mathcal{A}_{\text{related}}\} \) in offline inverted retrieval: $O(M*(N+K))$. (2) get all three parts of consultation value and aggregate the $\mathcal{O}_u^{s,c}$: $O(N*M) + O(M^2) + O(M^2)$ (3) consultation value alignment needs to compute CAI and use cascade encoder to compute $\mathcal{L}_{\text{VA}}$ and $\mathcal{L}_{\text{search}}$: $O(N'^2d)$, where $d$ is the encoding dimension, and $N'=\text{max}(M,N,K)$. 

Among all three parts, former two parts can be computed offline, and has lower complexity than $O(N'^2d)$, i.e. $O(M*(N+K)) + O(N*M) + O(M^2) + O(M^2)<O(N'^2) <O(N'^2d)$, so final time complexity of VAPS is $O(N'^2d)$.

\section{Appendix: Dataset License}
Following \citet{maps}, we provide details of dataset license used in this paper. 
The Amazon dataset (based on PersonalWAB~\citep{cai2024personalwab}) is released under the CC BY-NC 4.0 License. Its benchmark implementation, which is built upon the MIT-licensed tau-bench~\citep{yao2024tau}, incorporates significant modifications and enhancements tailored to the project's requirements. For the derived components, the implementation strictly adheres to and complies with the licensing terms of tau-bench.

\section{Appendix: Dataset Details}
\subsection{Dataset Repository}
We publicly disclose the dataset used in this paper at the following link: \url{https://anonymous.4open.science/r/VAPS-to-go}. Note that we only disclose the \textbf{Amazon} dataset, since \textbf{Commercial} dataset is currently not available for publicity because of policy and law restriction. Also, we give the dataset example as follows (same as \cite{maps}, due to we use the same datasets as theirs).

\begin{figure}[!t]
\begin{subfigure}[h]{0.5\textwidth}
    \centering
    \includegraphics[width=\textwidth]{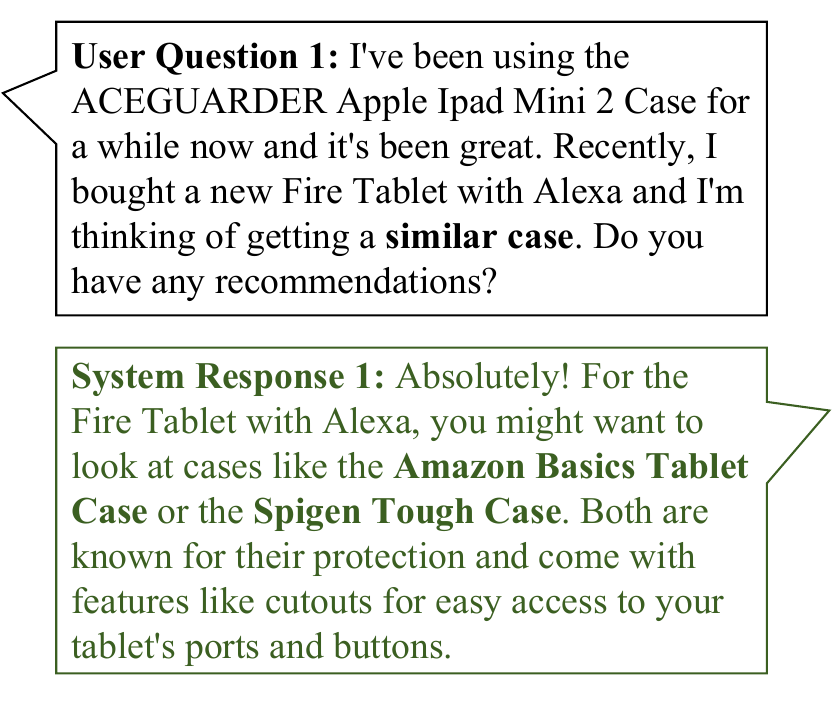}
\end{subfigure}
\begin{subfigure}[h]{0.5\textwidth}
    \centering
    \includegraphics[width=\textwidth]{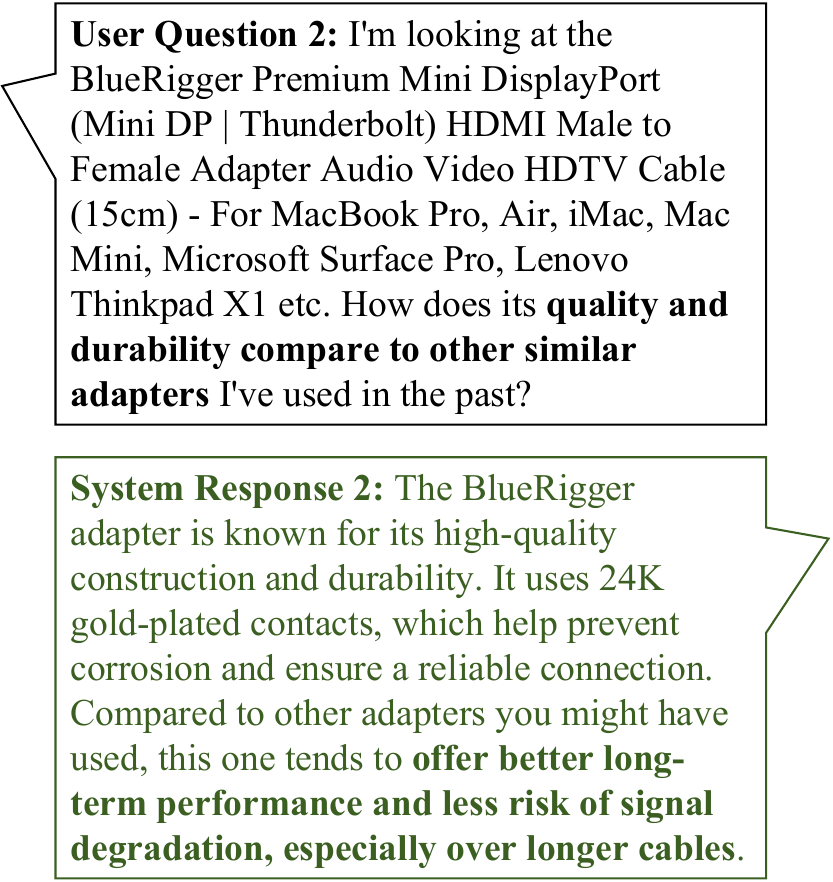}
\end{subfigure}
  \caption{Examples of consultations on the \textbf{Amazon} dataset.}
  \label{fig:app:amazon:1}
\end{figure}

\begin{figure}[!t]
\begin{subfigure}[h]{0.5\textwidth}
    \centering
    \includegraphics[width=\textwidth]{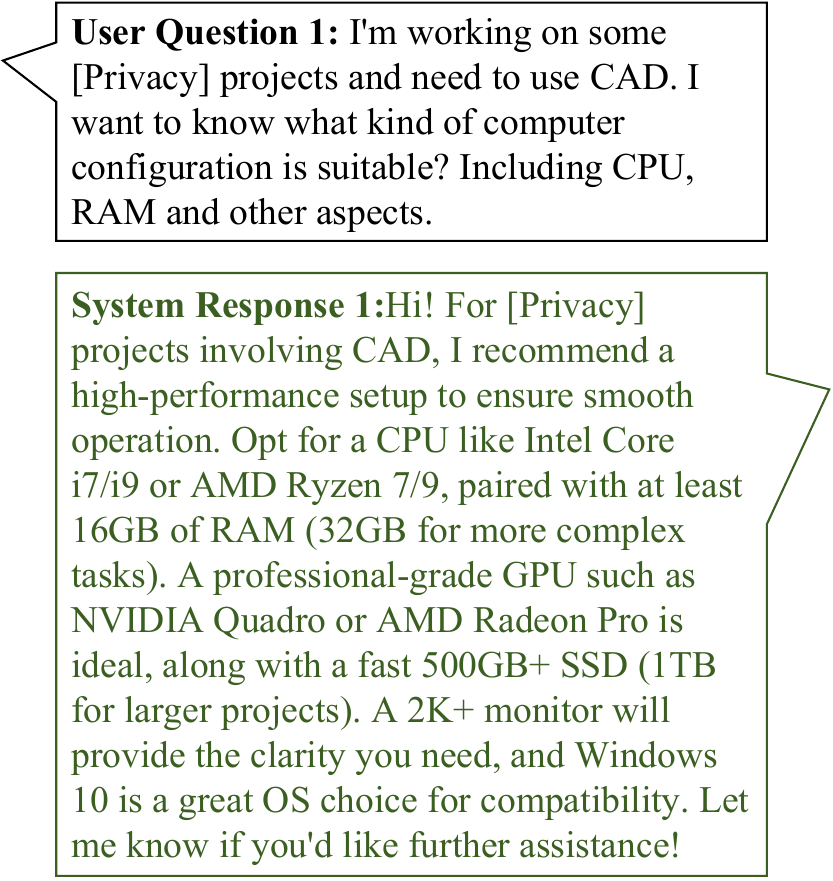}
\end{subfigure}
\begin{subfigure}[h]{0.5\textwidth}
    \centering
    \includegraphics[width=\textwidth]{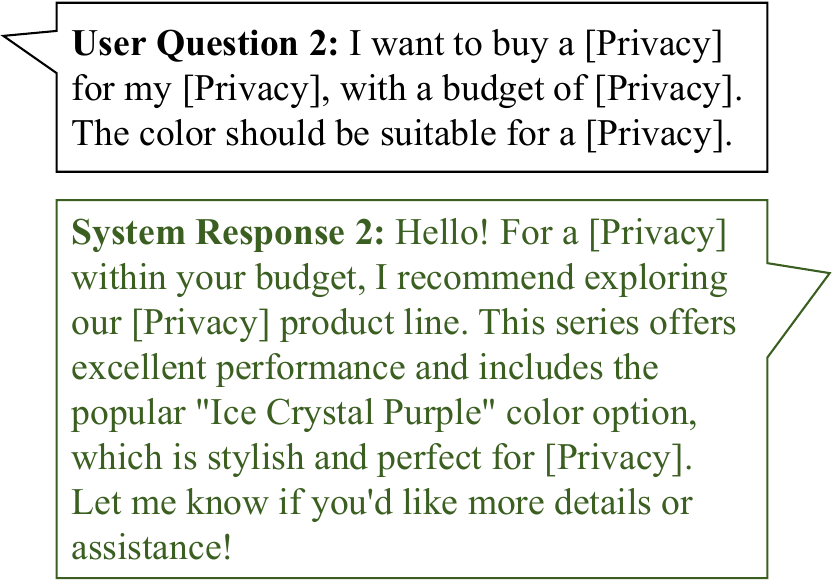}
\end{subfigure}
  \caption{Examples of consultations on the \textbf{Commercial} dataset.}
  \label{fig:app:commercial:1}
\end{figure}

\section{Appendix: Discussion}

The proposed Consultation Value Assessment Framework and VAPS model address critical gaps in consultation-enhanced personalized search by moving beyond semantic similarity to evaluate consultations through three novel dimensions: Scenario Scope, Posterior Action, and Time Decay Value. While these design choices are empirically justified, some aspects may invite scrutiny—particularly the heuristic nature of value functions, hyperparameter sensitivity, and dataset constraints. Here, we discuss the rationale behind these decisions and their implications.  

First, the time decay value ($\mathcal{O}^{s,c}_\text{time}$) employs an exponential decay to model diminishing consultation similarity over time. Although alternative formulations (e.g., linear or step-wise decay) could be considered, exponential decay that we adopted aligns with prior work on temporal dynamics in user behavior~\cite{rubin1996one} and effectively captures the intuition that recent interactions better reflect immediate intent and even recent but irrelevant consultations leave some impression on users with a certain value. The choice of $\alpha$ as a hyperparameter ensures simplicity, though adaptive decay rates (e.g., user-specific or activity-dependent) could further enhance personalization.  

Second, the scenario scope value ($\mathcal{O}^c_\text{scope}$) uses a necessary thresholding approach to filter out-of-domain consultation noise (e.g., politics discussions). Inverted indices provide a effective and practical way to identify in-scope terms, the threshold was selected empirically to balance precision and recall. A stricter threshold might exclude marginally relevant consultations, whereas a lenient one risks noise inclusion. Future work could explore dynamic thresholds based on consultation length or domain specificity.  

Third, the posterior action value ($\mathcal{O}^c_\text{action}$) quantifies consultation utility through subsequent user actions (e.g., clicks, purchases, searches). The bucketing of action frequencies (see App.~\ref{app:details:bucket})  is a efficient and stable method to mitigate long-tail distribution issues and ensures robust normalization, providing sufficient granularity without overfitting to sparse data~\cite{guo2017deepfm}. The relative weighting of actions reflects their inherent scarcity (e.g., purchases > clicks), a principle supported by consumer behavior studies~\cite{teubner2020only}.  

The aggregated value ($\mathcal{O}^{s,c}_u$) combines these dimensions via weighted summation, ensuring interpretability and mitigating the sparsity issue of value scores. Ablation studies (Sec.~\ref{sec:exp:ablation}) confirm that all components contribute to performance. The alignment loss $\mathcal{L}_\text{VA}$ further ensures that high-value consultation-action are captured by the model during training, though the temperature parameter \(\tau_1\) could be adapted per dataset to sharpen or soften attention supervision.  

Dataset limitations (e.g., sparsity in Commercial data) are acknowledged, yet results generalize well to the larger Amazon dataset. Since we consider an online personalized search scenario, LLM-based baselines are not included due to their excessive inference and training time costs. In fact, VAPS builds upon MAPS's MoAE (Mixture of Attention Experts) to integrate textual semantics from a frozen LLM embedding layer, ensuring efficiency while maintaining strong performance. Besides, VAPS's modular design (e.g., CAI) could integrate LLM-enhanced consultations in future work.  

In summary, while certain design choices (e.g., fixed decay rates, bucketing) are simplified for reproducibility, they are grounded in empirical evidence and previous works. The framework's flexibility allows for incremental refinements—such as dynamic thresholds or adaptive weighting—without undermining its core contributions. VAPS advances personalized search by assessing consultation value to users and aligning consultations with consumer actions, offering a scalable foundation for future enhancements.

\end{document}